\documentclass[%
 aapm,
 mph,%
 amsmath, amssymb,
preprint,%
]{revtex4-2}
\usepackage{graphicx}
\usepackage{dcolumn}
\usepackage{bm}

\usepackage[mathlines]{lineno}

\begin{document}

\title{Physics-constrained Bayesian inference of state functions in classical density-functional theory}

\author{Peter Yatsyshin$^*$\\ pyatsyshin@turing.ac.uk}
\affiliation{The Alan Turing Institute}

\author{Serafim Kalliadasis} 
\affiliation{Department of Chemical Engineering, Imperial College London}

\author{Andrew B. Duncan}
\affiliation{Department of Mathematics, Imperial College London; The Alan Turing Institute}

\date{\today}

\begin{abstract}

\section*{Abstract}
We develop a novel data-driven approach to the inverse problem of
classical statistical mechanics: given experimental data on the collective motion of
a classical many-body system, how does one characterise the free energy landscape of that system?
By combining non-parametric Bayesian inference with physically-motivated constraints, 
we develop an efficient learning algorithm which automates the construction of 
approximate free energy functionals. 
In contrast to optimisation-based machine learning approaches, which seek to minimise a cost function, 
the central idea of the proposed Bayesian inference is to propagate a set of prior assumptions 
through the model, derived from physical principles. The experimental data is used to probabilistically weigh the 
possible model predictions. This naturally leads to humanly interpretable algorithms with full
uncertainty quantification of predictions.
In our case, the output of the learning algorithm is a probability distribution over a family of free
energy functionals, consistent with the observed particle data. We find that surprisingly 
small data samples contain sufficient information 
for inferring highly accurate analytic expressions of the underlying 
free energy functionals, making our algorithm highly data efficient. 
We consider excluded volume particle interactions, which are ubiquitous in nature, whilst being highly challenging for modelling
in terms of free energy. To validate our approach we consider the paradigmatic case of 
one-dimensional fluid and develop inference 
algorithms for the canonical and grand-canonical statistical-mechanical ensembles.
Extensions to higher-dimensional systems are conceptually straightforward, whilst
standard coarse-graining techniques allow one to easily incorporate attractive interactions. 
\end{abstract}


\maketitle

\section{Introduction}
The past few years have seen an explosive development
of machine learning (ML) methods, which enabled dramatic 
enhancements across such diverse fields as pattern recognition~\cite{guyon1994}, natural language
processing~\cite{lakeS2015} and even DNA sequencing~\cite{alipanahiNB2015}.
It is now generally accepted that adoption of ML methods has the potential to
accelerate the development and enhance the quality of research across most scientific 
and engineering disciplines. Not
surprisingly, we are witnessing the emergence of a number of fields at the
intersection between ML and sciences-engineering: from quantum ML to
data-centric engineering. Despite its advantages and numerous success
stories, however, ML has yet to fulfil its full potential, especially in
long-standing fundamental problems.

One such classical problem comes from the field of statistical mechanics,
a branch of physics which aims to relate the observable macroscopic properties 
of matter with its underlying microscopic structure. Such properties as pressure, 
magnetisation or electric charge can all be determined by carefully averaging the small-scale 
interactions between the constituent particles of matter. The central property which facilitates 
such averaging is the one-body density function $\rho({\bf r})$, which can be thought of as 
the probability density function of finding a particle in the vicinity of the position-vector ${\bf r}$. 
Thus, a general method for obtaining $\rho({\bf r})$ of a given many-body system, poses a 
long-standing problem of fundamental importance to numerous scientific and engineering fields.

Even when the system particles interact via simple potentials, computing $\rho({\bf r})$ 
exactly is computationally intractable, due to inter-particle correlations. For example, 
application of the Liouville theorem leads to a hierarchy of
density correlation functions and requires simplifying closure assumptions 
to make the resulting system of equations computable. A way out is offered by the density-functional 
theory (DFT), which is based on the mathematical 
fact that the free energy of a many-body system is a functional of its density $\rho({\bf r})$ and attains its
minimum at the density of the system at equilibrium.
In the subject literature the acronym DFT usually refers to one of the two generic classes of models: 
(i) quantum DFT, which deals with the exchange-correlation energy of quantised many-body systems, 
particularly electron gas~\cite{HohenbergKohnPhysRev1964, levyPotNAoS1979};
and (ii) classical DFT (cDFT), which applies to many-body systems with 
classical interactions such as common liquids, electrolytes, salts 
etc. and works with the Helmholtz free energy functional 
$F[\rho]$~\cite{EvansAdvPhys1979, WuLiAnnuRevPhysChem2007, LutskoAdv.Chem.Phys.2010}.  

The focus here is on the classical world. Ab initio quantum DFT calculations
promise to realise the full potential of statistical mechanics. However, they
are computationally forbidding for beyond-the-molecular-scale systems, despite
drastic improvements in computational power. At the same time, classical
statistical mechanics and cDFT are not totally disconnected from the quantum
world as they subsume many quantum effects inside the particle interaction potentials.
Not surprisingly, cDFT is a generic and widely used statistical mechanical 
framework for numerical and mathematical scrutiny.  

Unfortunately, the payoff reaped by the DFT formulation of the many-body problem as the minimisation of 
the free energy is lessened by the fact that the exchange correlation energy in quantum DFT
and the excess-over-ideal Helmholtz free energy in cDFT are not known exactly.
This necessitates the development of methods to approximate these terms,
which form the bulk of modern statistical-mechanical literature.

A popular intuitive and practical method for constructing DFT approximations is based on 
coarse-graining the intermolecular interactions~\cite{WuLiAnnuRevPhysChem2007}. 
Essentially, coarse-graining splits the unknown $F[\rho]$ of the system
into appropriate reference and perturbation parts, treating them
separately. The reference system describes the dominant interaction, which in many 
systems corresponds to the repulsive part of the full intermolecular potential.
For example, interactions in a Lennard-Jones fluid consist of short-range
repulsions, caused by the overlap of the electron orbitals and the Pauli
exclusion principle, and comparatively weaker long-range attractions, caused
by the dipole-dipole electromagnetic interaction. A coarse-grained approximate $F[\rho]$
is then given by the free energy of a hard sphere fluid with an added
mean-field attractive term. Coarse-graining can be viewed as an extension of the
Born-Oppenheimer approximation, whereby intermolecular
effects occurring on different spatiotemporal scales are decoupled and
accounted for separately.

The majority of realistic many-body systems can, in principle, be described by similar
techniques. Indeed many studies have highlighted the potential utility of
coarse-grained cDFTs in practical applications, including phase transitions,
interfacial phenomena, colloidal and polymer fluids, surfactants, liquid
crystals, crystalline solids, glasses and the rapidly growing fields of
microfluidics~\cite{WuLiAnnuRevPhysChem2007, lutskoSA2019}. Yet, despite considerable
efforts to bring cDFT to the applications
domain~\cite{seguraMP2001,tripathiTJoCP2003}, existing cDFT approximations are mainly
limited to highly idealised systems. Thus, present-day cDFT is still far from 
becoming an instrument of widespread practical utility and obtaining the status of 
a computational go-to framework, on a par with molecular dynamics (MD) and computational fluid dynamics \cite{chiavazzoNC2014}.

As mentioned above, one important caveat is to adequately capture the reference system of purely repulsive
hard particles of a given shape. In applications, fluid particles interact via complex
potentials which makes the construction of DFTs in each case extremely
difficult. There is a clear need for an algorithmic hands-off method for obtaining
accurate and robust DFT functionals of purely repulsive systems. Coming to our rescue, modern 
statistical inference frameworks offer the principle means to develop just such a method.

While there is a growing number of ML works in quantum DFT,
applications to classical many-body systems are still extremely rare. In
Ref.~\onlinecite{yousefzadinobakhtAA2020}, a Bayesian approach was developed to fit
the drift of a Langevin equation, describing oscillations of an atom in a
lattice. The posterior was sampled with sequential Monte-Carlo, which
accommodates large datasets, but has the downside of being suitable only for
parametric models with a small number parameters. The four parameters of a
postulated ansatz for the inter-particle potential were fitted using
simulation data. Unlike Ref.~\onlinecite{yousefzadinobakhtAA2020}, where the number
of parameters is fixed, our statistical model for $F[\rho]$ is non-parametric,
thus possessing a high-level of flexibility. Our method is also technically
more advanced in that it employs
adjoint differentiation to evaluate the solution gradient on every step of
the sampler. In Ref.~\onlinecite{linJCP2020b}, the choice of free energy terms is
formulated as a classification problem employing a neural network (NN). The
training is done by minimising the regularised Euclidean distance between
the trained particle distribution and another distribution, obtained by
averaging the simulation data. The choices of the NN architecture and the
loss function are highly empirical. Additionally, raw simulation data
requires costly post-processing, leading to information loss and compromising
the accuracy of the predictions. Lastly, no uncertainty estimates are
provided in Ref.~\onlinecite{linJCP2020b} for the trained functional, raising
applicability concerns. In contrast, uncertainty propagation and
quantification are central to our proposed method. Our algorithm keeps down 
empirical choices required of the user and trains on raw particle data. In
ML quantum DFT works NNs are quite popular\cite{nagainCM2020, chandrasekarannCM2019}. 
While deep NNs are well-suited for approximating
functions of a high-dimensional arguments and can benefit from automatic
differentiation, constructing NN architecture and training are essentially
black-box. NNs suffer from interpretability issues, they do not natively provide
uncertainty quantification and require large data sets for training. In contrast, 
our work focuses on physics-constrained learning, requires small data sets and takes
full advantage of the Bayesian paradigm, enabling interpretability and
quantification of uncertainty. Lastly, the generality and versatility of our
approach makes it transferable across statistical mechanics and beyond,
including quantum DFT applications.

\section{Statistical model of a hard-body free energy functional}
The exact cDFT functional $F[\rho]$ is known only for a one-dimensional (1D)
fluid of hard rods (HR), constrained to a line \cite{PercusJStatPhys1976}.
Approximate DFTs have been developed for a handful of relatively simple
idealised systems, such as hard spheres, hard discs or parallel hard cubes
\cite{TarazonaCuestaEtAlTheoryandSimulationsofHard-SphereFluidsandRelatedSystems.LectureNotesinPhysics7532008}.
The aim of the present work is to learn $F[\rho]$ of a hard-particle fluid,
using particle trajectories obtained from small-scale simulations. Since our
goal is methodological, we consider the simplest possible case of 1D HR on a
line. This system provides an optimal starting point for two reasons. First,
access to the exact $\rho({\bf r})$ allows us to easily benchmark inference
against ground truth. Second, data generation is cheap in 1D, which in turn
facilitates convergence studies and comparison to brute-force statistical
inference. Application of our approach to more complex systems is
conceptually straightforward, but would require an increased computational
effort. It should be noted that in higher than one dimension fluids can
undergo phase transitions. Still, coarse-graining techniques can be used for
such fluids: by first treating the reference system with purely repulsive
interactions, the possibilities of liquid-gas coexistence and criticality are
eliminated. The cDFT of an attractive-repulsive fluid can then be obtained
by, e.g., adding a simple mean-field attractive term to the reference cDFT.
The general strategy of splitting interactions leads to good approximations
in many cases potentially of practical interest
\cite{WuLiAnnuRevPhysChem2007, LikosPhys.Rep.2001}. Clearly, hard-particle
fluids can still undergo freezing, and this possibility can be built into
$F[\rho]$ by using appropriate trial functionals with singular terms.
However, in many liquid-state problems the temperature is sufficiently high
that the fluid is not frozen. In such cases a coarse-grained cDFT can capture
a wide spectrum of phenomena, e.g. surface-phase transitions during
adsorption \cite{YatsyshinParryEtAlJPhysCondensMatter2016}, even when the
reference functional does not properly describe the limiting case of
freezing.

\subsection{Direct and inverse problems of statistical mechanics}
\label{SecDP} The direct problem of equilibrium statistical mechanics in the
cDFT formulation can be stated as follows. Obtain the probability-density
function $\rho({\bf r})$ over the positions of $N$ interacting particles
moving in an external field $V({\bf r})$ by minimising the \emph{given}
free-energy functional $F[\rho]+\int{\rho({\bf r})V({\bf r})d{\bf r}}$. In
additions, the number of particles may be constrained, $\int{\rho({\bf r})
d{\bf r}} = N$. Observe that the functional $F[\rho]$ above is independent of
$V({\bf r})$. A singular $V({\bf r})$ can describe the geometric confines of
the particles. Thus, knowing $F[\rho]$ allows us, in principle, to compute
the collective statistics of the system in any spatially confined setting, as
well as in the bulk. Formally, we can cast the minimisation problem in terms
of a Lagrangian $\Omega[\rho]$, introducing a new variable $\mu$ as the dual
of $\rho$:
\begin{align}
\label{OM}
\Omega[\rho]=F[\rho]+\int{\rho({\bf r})\left(V({\bf r})-\mu\right)d{\bf r}}.
\end{align}
In equilibrium, the minimum of $\Omega[\rho]$ can be formally obtained from
the system's Hamiltonian, by computing the grand-canonical partition function
\cite{AllenTildesley1989}. Thus, we can get physically meaningful results by
applying just the first part of the minimax principle, minimising
$\Omega[\rho]$ at a given $\mu$. Fixing $\mu$ instead of $N$ is equivalent to
considering an open system, where $N$ fluctuates around its average
$\mu$-dependent value $\langle N_\mu\rangle$. In this case, we say that the
system is connected to a particle reservoir, held at the chemical potential
$\mu$. Considering systems at fixed $N$ is known as the canonical ensemble,
whereas fixing $\mu$ instead is the grand-canonical ensemble. In large
systems both ensembles are equivalent, but systems of a few particles may
exhibit differences between ensembles \cite{LebowitzPercusEtAlPhysRev1967}.
It should be noted that the form of \eqref{OM} is the same in both ensembles,
and the differences are subsumed by the definition and interpretation of
$F[\rho]$. Traditionally, a grand-canonical ensemble is implied with cDFT,
but we will demonstarte that it is possible to statistically infer both,
canonical and grand-canonical representations from the relevant particle
data.

The inverse problem of statistical mechanics can be formulated as finding the
\emph{unknown} $F[\rho]$, using a number of observations of instantaneous
coordinates of the system's particles. The data can be obtained using
Monte-Carlo or MD simulations in the relevant ensemble
\cite{AllenTildesley1989}. We seek to compute a probability distribution over
the free energy functionals, consistent with the data. This approach is
markedly different from traditional analytic modelling, which aims to
construct a single approximation for $F[\rho]$, valid under some idealised
conditions.

\subsection{Free-Energy Model}
In classical systems, interactions between the particles are described by the
excess-over-ideal part, $F_{ex}[\rho]$, of the full free-energy functional
$F[\rho]$:
\begin{align}
\label{F}
&F[\rho] = \beta^{-1}\int{ \rho({\bf r})\left(\ln{\lambda^3\rho({\bf r})}-1\right)d{\bf r}}+F_{ex}[\rho],
\end{align}
where $\lambda$ is the thermal wavelength, which includes the contribution
from the Maxwell distribution of particle velocities, and $\beta$ is the
inverse temperature. We notice that changing $\lambda$ is equivalent to
changing $\mu$ in \eqref{OM}, which allows us to set $\lambda=1$ without loss
of generality. In the grand-canonical ensemble, after $\rho({\bf r})$ is
obtained by minimising \eqref{OM}, the correlation-function hierarchy can be
recovered by computing the inverses of the functional derivatives of
$F_{ex}[\rho]$ at $\rho({\bf r})$. Thus, cDFT is nothing but a convenient
formulation of statistical mechanics. When particle interactions are pairwise
and given by the potential $\phi(r)$,  the following expansion is valid for
$F_{ex}[\rho]$ \cite{LutskoAdv.Chem.Phys.2010}:
\begin{align}
\label{virial}
\beta F_{ex}[\rho] =& -\frac{1}{2}\int{\rho({\bf r_1})d{\bf r_1}}\int{\rho({\bf r_2}) f({r_{12}})d{\bf r_2}}\notag\\
&+\frac{1}{6}\int{\rho({\bf r_1})d{\bf r_1}}\int{\rho({\bf r_2})d{\bf r_2}}\int{\rho({\bf r_3}) f({r_{12}})f({ r_{23}})f({r_{13}})d{\bf r_3}}
+\mathcal{O}(\rho^4),
\end{align}
where ${r_{ij}}=\left.|{\bf r_i}-{\bf r}_j\right.|$ and
$f(r)=\exp{(-\phi(r))}-1$ is the Mayer function. Interactions in a
hard-particle fluid are purely repulsive, caused by volume exclusion and the
fact that particles have finite sizes and impenetrable cores. For such fluids
the Meyer function equals -1 in the spatial regions where the particles
overlap and zero otherwise. Thus, $f(r)$ can be expressed as a weighted sum
of convolutions of the so-called geometric \emph{fundamental measures} --
window-functions $\{\omega_{i}({\bf r})\}$, which characterize particle
geometry in terms of volume, surface area, Gaussian and deviatoric
curvatures, etc. \cite{wittmannTJoCP2014}. Using this fact, we can cast the
low-density asymptote of \eqref{virial} in terms of the weighted densities
$n_i({\bf r})$, given by the convolutions of $\rho({\bf r})$ with each
$\omega_{i}({\bf r})$:
\begin{align}
\label{limit}
{\beta F_{ex}[\rho]}\underset{\rho\to0}{\sim} -\int{\left.\sum\limits_{i,j}n_i({\bf r})n_j({\bf r})\right.d{\bf r}},\quad\mathrm{where}\quad n_i({\bf r})=\omega_i* \rho\equiv\int{\omega_i({\bf r}+{\bf t})\rho({\bf t})d{\bf t}}.
\end{align}
The number of terms in the sum above depends on the number of non-zero
fundamental measures and is determined by the particle shape and
dimensionality. For example, a sphere of radius $R$ can be described by
$n_i({\bf r})$, obtained from two scalar-valued functions
$\Theta(R-\left|{\bf r}\right|)$ and $\delta(R-\left|{\bf r}\right|)$,
yielding the sphere volume and surface area, and one vector-valued function
${\bf r} \delta(R-\left|{\bf r}\right|)/\left|{\bf r}\right|$, yielding the
mean curvature
\cite{TarazonaCuestaEtAlTheoryandSimulationsofHard-SphereFluidsandRelatedSystems.LectureNotesinPhysics7532008}.
Here $\Theta(x)$ and $\delta(x)$ are the Heaviside function and the Dirac
delta-function. In the case of 1D HR fluid with HR of width $2R$, ${\bf
r}\equiv x$ and there are only two fundamental measures, the volume and
surface ones:  $\omega_v(x)=\Theta(R-|x|)$ and $\omega_s(x)=\delta(R-|x|)/2$.
These give rise to the respective weighted densities $\eta(x)$, and $n_0(x)$:
\begin{align}
\label{FM}
\eta(x)=\int\limits_{x-R}^{x+R}\rho(t){dt},\quad n_0(x)=\frac{\rho(x-R)+\rho(x+R)}{2}.
\end{align}

The fact that the asymptote of $F_{ex}[\rho]$ in \eqref{limit} is a local
functional of $\{n_i\}$ suggests to approximate $F_{ex}[\rho]$ in the form of
functions of $\{n_i\}$. This simple intuition also forms the physical basis
of our inference framework:
\begin{align}
\label{Fex}
F_{ex}[\rho]=\beta^{-1}\int{\Phi(n_1({\bf r}), n_2({\bf r}), \dots)d{\bf r}},
\end{align}
where $\Phi(\{n_i\})\equiv\Phi(n_1({\bf r}), n_2({\bf r}), \dots)$ is a
multivariate function of $\{n_i\}$. Observe that the 2nd and higher terms of
\eqref{virial} cannot be directly expressed as functions of the weighted
densities. Thus, \eqref{Fex} is indeed just an approximation for
extrapolating the asymptote in \eqref{limit} to higher $\rho$. Furthermore,
the function $\Phi(\{n_i\})$ is not unique, e.g., any function which
integrates to zero can be added to it. Over the years, many sophisticated
theories of increasing complexity were proposed for $\Phi(\{n_i\})$. These
gave rise to a plethora of approximate cDFTs, collectively known as the
Fundamental Measure Theory (FMT)
\cite{TarazonaCuestaEtAlTheoryandSimulationsofHard-SphereFluidsandRelatedSystems.LectureNotesinPhysics7532008}.
Yet, even in the case of simple hard-sphere fluids, a universally acceptable
approximate $F_{ex}[\rho]$ remains elusive: some functionals fail to recover
the thermodynamic equation of state, others diverge when particle motion is
restricted to low-dimensional manifolds, others still fail to adequately
capture the freezing of hard spheres. In engineering applications, a
hard-sphere cDFT is commonly used as a reference part of a more complex
coarse-grained functional of an attractive-repulsive fluid, such as a
Lennard-Jones fluid or a polymer chain~\cite{zhangMP2020, WuLiAnnuRevPhysChem2007}. Typically, there also
is a well-defined range of temperatures and pressures of interest. For such
restricted regimes, in most practical cases one can select a satisfactory
hard-sphere approximation.

Apart from hard spheres, FMTs are obtained only for a handful of simple
molecular shapes
\cite{TarazonaCuestaEtAlTheoryandSimulationsofHard-SphereFluidsandRelatedSystems.LectureNotesinPhysics7532008}.
However, biology, colloidal and polymer physics, fluid particles often have
complex non-spherical shapes. As a result, analytic construction of
approximate functionals for each particular problem is extremely difficult,
if not impossible. In what follows, we borrow the common aspect of the best
existing FMTs, expressed by \eqref{Fex}, and develop a Bayesian approach to
the inference of $\Phi(\{n_i\})$ from the simulated particle trajectories. As
mentioned above, for methodological simplicity we consider the paradigmatic
case of a 1D HR system.

\section{Bayesian inference of the grand-canonical density functional}
A system of HR inside a pore of width $L$ is sketched at the top of
Fig.~\ref{FigOne}(a). When the fluid is held at the chemical potential $\mu$,
its collective behaviour in the grand-canonical ensemble can be simulated,
yielding the expected number of particles $\langle N_\mu\rangle$ in the pore
and a set of $M$ instantaneous particle positions $\{y_i\}_{i=1}^{M}$. These
form our fixed-$\mu$ training data set $ \mathcal{D}_\mu$:
 \begin{equation}
 \label{DataMu}
 \mathcal{D}_\mu=\left(\mu,\{y_i\}_{i=1}^{M},\langle N_\mu\rangle\right).
 \end{equation}
The simulation algorithm is described in Sec.~Methods. When $M$ is
sufficiently large, the normalised histogram of $\{y_i\}_{i=1}^{M}$ should
approximate the DFT density profile $\rho(x)$, which minimizes $\Omega[\rho]$
in \eqref{OM}. We assume that $F_{ex}[\rho]$ is given by \eqref{Fex} with an
unknown function $\Phi(n_0, \eta)$ of two weighted densities, given in
\eqref{FM}. Thus, $\rho(x)$ solves the Euler-Lagrange equation:
\begin{align}
\label{EL}
\ln{\rho(x)}+\left(\omega_v *\frac{\partial\Phi}{\partial\eta(x)} + \omega_s*\frac{\partial\Phi}{\partial n_0(x)}\right)-\beta\mu = 0,\quad \mbox{subject to} \int\limits_{-L/2}^{L/2}{\rho(x)}dx = \langle N_\mu\rangle.
\end{align}

We adopt a straightforward and fairly general polynomial form of
$\Phi(n_0,\eta)$ in terms of parameters $Q$:
\begin{equation}
\label{Phi}
\Phi(n_0,\eta)\equiv\Phi(n_0,\eta\mid Q)=\left(a_{N_1} n_0(x)^{N_1}+a_{N_1-1} n_0(x)^{N_1-1} + a_0\right) \left(b_{N_2} \eta(x)^{N_2}+\dots b_{0}\right),
\end{equation}
where $Q=\left(a_{N_1}, \dots a_0, b_{N_2},\dots b_0\right)^T$ has $N_Q=N_1+N_2+2$ elements. Observe that \eqref{Phi} provides a highly flexible model, capable of representing a broad class of smooth functions. To avoid the equivalence between $\Phi(n_0,\eta\mid Q)$ and $\Phi(n_0,\eta\mid -Q)$, we constrain $a_0$ to be non-negative. During training we will be solving \eqref{EL} numerically for randomly drawn $Q$. For stability we use a simple Newton scheme and a trapezium rule for quadratures. Our goal is to find the distribution $P(Q)$, which in turn induces two other distributions: one over the free energy functionals $F[\rho\mid Q]$, via the term $\Phi(n_0,\eta\mid Q)$ and Eqs. \ref{Fex} and \ref{F}, and another one over the densities $\rho(x\mid Q)$, via \eqref{EL}. To characterise $P(Q)$ by a single value, one can compute the expectation $\rho_E(x) = \int \rho(x\mid Q) d P(Q)$, or alternatively, the maximum a-posteriori estimator (MAP), $\rho_{\mathrm{MAP}}(x)=\rho(x\mid \mathrm{argmax } P(Q))$.

As mentioned earlier, the free-energy functional of HR in the grand-canonical
ensemble is known exactly. It is given by the expression
$\Phi_X(n_0,\eta)=-n_0\log{(1-\eta)}$ \cite{PercusJStatPhys1976}. Thus, the
ground truth for the grand-canonical inference is given by
$\rho(x)\equiv\rho_{X\mid\mu}(x)$, which solves \eqref{EL} with
$\Phi\equiv\Phi_X(n_0,\eta)$. Since $\Phi(n_0, \eta)$ is not unique, we do
not expect to infer $\Phi_X(n_0,\eta)$ precisely.

\subsection{Inference Procedure}
Bayesian prior should characterise $Q$ it in the absence of training data.
Clearly, not every $Q$ yields an admissible free-energy functional. Hence, we
choose a Gaussian prior $\mathcal{N}(Q\mid \bar{Q},\Sigma_Q)$ with mean
$\bar{Q}=0$ and a diagonal covariance matrix $\Sigma_Q$. With this prior on
$Q$, $\Phi(\eta,n_0)$ is a Gaussian random field with polynomial features in
$\eta$ and $n_0$. The prior variances on $Q$ are chosen to constrain the
components of $Q$ to be sufficiently close to zero. The expression for
Bayesian likelihood follows from the physical interpretation of $\rho(x\mid
Q)$ as the probability-density function:
\begin{equation}
\label{likelihoodMu}
P(\mathcal{D}_\mu\mid Q) = \prod\limits_{i=1}^{M}\rho(y_i\mid Q).
\end{equation}

The posterior distribution over $Q$ follows from the Bayes rule, $P(Q \mid
\mathcal{D}_\mu)\propto \mathcal{N}(Q\mid \bar{Q},\Sigma_Q)
P(\mathcal{D}_\mu\mid Q)$, and yields predictive posterior distributions over
$F[\rho\mid Q]$ and $\rho(x\mid Q)$. Since $P(Q \mid \mathcal{D}_\mu)$ is not
analytically tractable and known up to a normalising constant, approximate
methods of inference must be considered. A popular approach is to use Markov
chain Monte-Carlo to construct a Markov chain with samples asymptotically
distributed according to $P(Q \mid \mathcal{D}_\mu)$. Here we implement a
Hamiltonian Monte-Carlo (HMC) algorithm to generate samples from the
posterior \cite{brooks2011}. At every iteration of the chain HMC needs the
gradient of log-posterior, $\nabla_Q\log{P(Q\mid \mathcal{D}_\mu)}$. This, in
turn, requires computing the Jacobian of the numerical solution to \eqref{EL}
with respect to $Q$, $\nabla_Q\rho(x\mid Q)$. It is easy to see that a direct
calculation of this Jacobian requires solving \eqref{EL} $N_Q+1$ times. The
fact that this calculation must be done at every iteration of the Markov
chain may render $\nabla_Q\log{P(Q\mid \mathcal{D}_\mu)}$ computationally
intractable even for moderate $N_Q$. Significant improvement in the
computation of $\nabla_Q\rho(x\mid Q)$ can be achieved by using adjoint
differentiation methods, which relate the numerical solution of \eqref{EL}
with its Jacobian via a linear system. The expressions for log-posterior and
its gradient are provided in Sec.~Methods. To simplify notation, we drop the
bar-notation for conditional probabilities.

\begin{figure}
\centering
\includegraphics[scale=1]{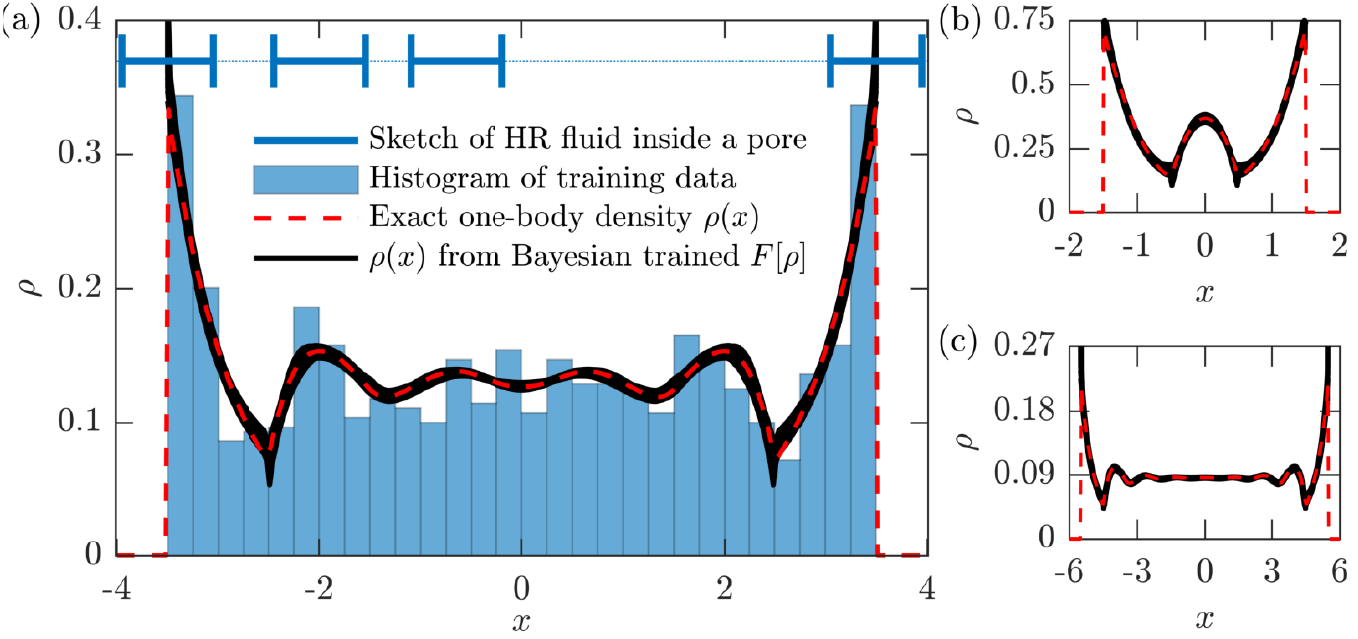}
\caption{Illustration of the trained HR functional. Simulated HRs have width $2R=1$ and interact via elastic collisions.
They are confined to a pore of width $L=8$ and are held at chemical potential $\mu=2$, so that $\langle N_{\mu}\rangle=4.6$.
The histogram in (a) shows the training dataset from \eqref{DataMu} with $M=1000$ simulated HR coordinates, used to train $F[\rho]$ with $N_1=N_2=5$ in \eqref{Phi}.
The black ``curve'' shows 200 profiles $\rho(x)$, obtained from samples of the trained DFT functional.
The spread of these profiles characterises the uncertainty of the Bayesian scheme about
$F[\rho]$. (b) and (c) show 200 samples in pores with $L=4$ and $L=12$, obtained from the same functional as (a). In (a)-(c),
the dashed red curve shows the ground truth in terms of exact distribution $\rho_{X\mid\mu}(x)$.}
\label{FigOne}
\end{figure}

After tuning the HMC step-size and burn-in parameters to ensure that the
output is stationary and sufficiently fast mixing, we generate samples using
a sufficiently long run of $4$ independent chains. The empirical predictive
distribution for $F[\rho]$ is then obtained from these samples analytically
via Eqs.~(\ref{Phi}) and (\ref{Fex}), and can be viewed as a distribution
over free energies, consistent with the observed simulation data. We
illustrate this in Fig.~\ref{FigOne}, where we train the grand-canonical
$F[\rho]$ at $\mu=2$ and $L=8$. Figure~\ref{FigOne}(a) shows the histogram of
the training data, 200 densities $\rho(x)$ minimising samples of $F[\rho]$
(black), and the ground truth, given by the exact distribution
$\rho_{X\mid\mu}(x)$ (dashed red). Observe that the inferred $F[\rho]$
represents the ground truth well, even though the training set histogram is
rather coarse and does not visibly approximate $\rho_{X\mid\mu}(x)$. This
attests to the ability of the physics-informed Bayesian method to combine
essential physical features with the data to achieve high efficiency of
inference. Figures~\ref{FigOne} (b) and (c) show the predictions of the same
$F[\rho]$ about the fluid in pores with out-of-sample $L=2$ and $L=12$. The
superimposed $\rho_{X\mid\mu}(x)$ again attests to the high quality of
inference. The spread of the prediction curves $\rho(x)$ is indicative of the
standard deviation and captures the local uncertainty. This seems largest
around turning points of the profiles. The uncertainty can be reduced by
increasing the size of the training dataset. For the wide pore in
Fig.~\ref{FigOne}(c) the effects of the side walls are lost in the pore
center, and the fluid near the pore center should behave like bulk fluid. The
fact that the correct plateau of $\rho_{X\mid\mu}(x)$ is reproduced by the
trained $F[\rho]$ means that the trained functional correctly captures the
physics of the bulk fluid and its thermodynamic equation of state. This
result is quite remarkable considering the fact that we were training on the
data of a highly confined fluid, represented by the histogram in
Fig.~\ref{FigOne}(a).

\subsection{A Gaussian random field model for chemical potential}
We cannot expect the model used in Fig.~\ref{FigOne} to provide good
predictions for $\mu$ outside of the training set, which currently includes
only a single $\mu$-point. To achieve generalisation with $\mu$ we must
extend the learning procedure in two ways: (i) by providing the training data
for multiple values of $\mu$ and. (ii) extending the model to be
$\mu$-dependent. At first glance, (ii) may seem inconsistent with \eqref{F},
where $F[\rho]$ does not explicitly depend on $\mu$. However, a more flexible
inference model may help us counteract the limitations of the finiteness of
the training sets and the finite dimensionality of $Q$. Certainly, in the
limit of $N_Q\to\infty$ and infinitely large training set $\mathcal{D}$, any
built-in $\mu$-dependence of $\Phi$ must disappear as the ground-truth
functional is recovered. On the other hand, when $\mathcal{D}$ and $N_Q$ are
finite, it is worthwhile to test the performance of the $\mu$-dependent
inference model on interpolation and extrapolation to out-of-sample
$\mu$-points.

We generalise $\Phi(n_0,\eta\mid Q)$ to $\Phi(n_0,\eta\mid Q(\mu))$ by
representing each element of $Q$ as a polynomial of degree $M$. The new
parameter set is represented by the $N_Q\times(M+1)$ matrix $A$ of polynomial
coefficients:
\begin{align}
\label{PhiAlpha}
\Phi(n_0,\eta|\mu,\alpha)\equiv\Phi(n_0,\eta\mid Q(\mu\mid\alpha)),\quad Q(\mu\mid \alpha) = A\left(\mu^M, \mu^{M-1} \dots1\right)^T,
\end{align}
where $\alpha=\left(\alpha_1, \dots \alpha_{N_\alpha}\right)^T$, $N_\alpha=N_Q(M+1)$, is the (row-wise) flattened matrix $A$. The training data set $\mathcal{D}$ and the likelihood function for this extended model become:
\begin{align}
&\mathcal{D} = \{\mathcal{D}_{\mu_n}\}_{n=1}^K\equiv \left\{\left(\mu_n,\{y_{i\mid n}\}_{i=1}^{M_n},\langle N_n\rangle\right)\right\}_{n=1}^K,\label{Data}\\
&P(\mathcal{D}\mid\alpha)= \prod\limits_{n=1}^K\prod\limits_{i=1}^{M_{n}}\rho(y_{i\mid n}\mid\mu_n,\alpha)  ,\label{likelihood}
\end{align}
where $M_n\equiv M_{\mu_n}$, $y_{i\mid n}$ is the $i$-th simulated particle
coordinate at $\mu_n$, $\langle N_n\rangle\equiv\langle N_{\mu_n}\rangle$,
and $\rho(x\mid\mu_n,\alpha)$ is the solution of \eqref{EL} at $\mu=\mu_n$
and $\Phi$ given by Eqs.~(\ref{Phi}) and (\ref{PhiAlpha}). Now a joint
Gaussian prior on the coefficients $\alpha$ induces a Gaussian random field
prior on the space of functions of $n_0$, $\eta$ and $\mu$.

\begin{figure}
\centering
\includegraphics[scale=1]{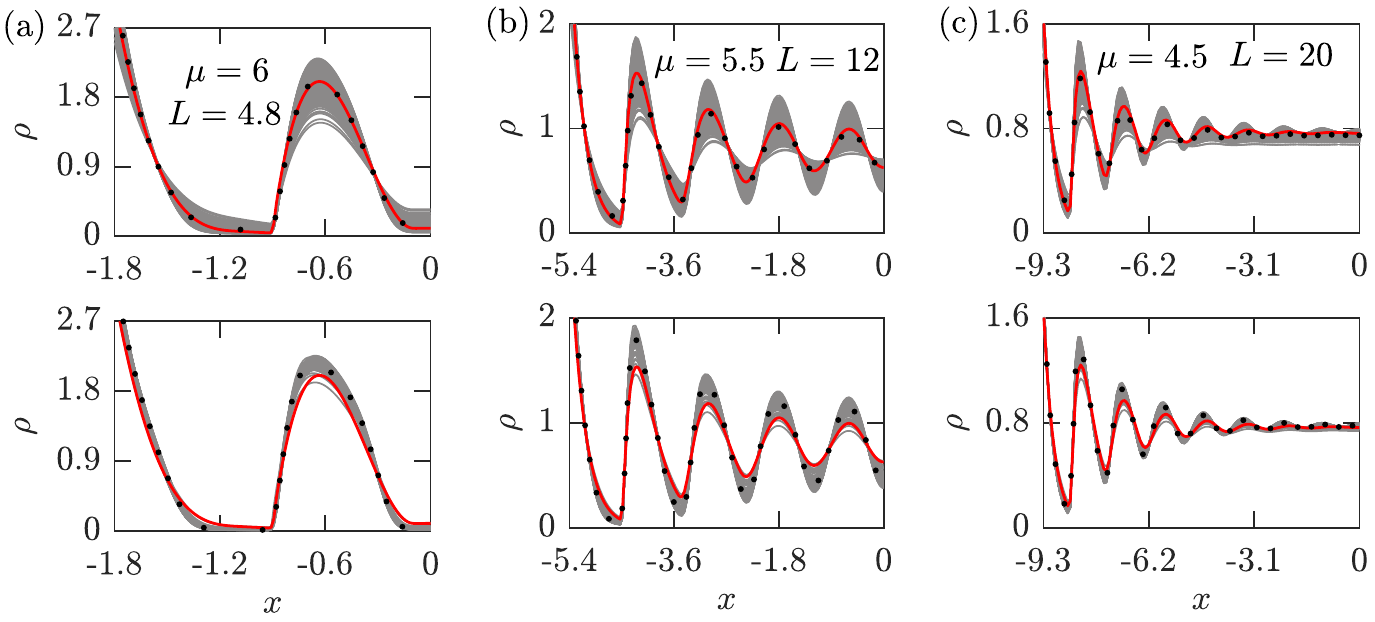}
\caption{Generalisation of the trained functionals with chemical potential $\mu$, as expressed by Eqs.~(\ref{PhiAlpha})--(\ref{likelihood}).
Two functionals, both with $N_1=3$ and $N_2=8$ (but different $M$) are trained at $K=8$ integer values of $\mu=-2,\dots5$, using $M_n=10^4$ particle coordinates
per $\mu$-point. Top and bottom panels in (a)-(c) correspond to $M=0$ (no $\mu$ dependence) and $M=1$ (linear $\mu$ dependence). Depicted density profiles minimise \eqref{OM}
at the specified $\mu$ and $L$, and due to symmetry are shown for $-L/2\leq x\leq0$. The posterior spread is illustrated by 400 samples (grey).
The MAP estimators (black dots) lie close to the ground truth, $\rho_{X\mid\mu}(x)$ (red). Notice that the linear in $\mu$ model has less uncertainty,
but the $\mu$-independent model has better MAP estimators, particularly in (a) and (b), where $\mu$ extrapolates from the training set. This exemplifies over-fitting.}
\label{FigTwo}
\end{figure}

As before, Eqs.~(\ref{PhiAlpha})--(\ref{likelihood}) define a posterior
distribution over $\alpha$ for the given particle data, and we obtain the
corresponding distribution over $F[\rho]$ from that posterior. But we can now
characterise the fluid for a broad range of $\mu$, including very dilute
(small $\mu$) and highly structured (large $\mu$) fluid configurations, using
the same posterior. As an example, consider a training dataset with $K=8$
integer $\mu$-points, $\mu=-2\dots 5$, and $M_n=10^4$ particle coordinates
per $\mu$-point, drawn from the grand-canonical simulation in a pore of width
$L=8$. We use this data to train two functionals: a $\mu$-independent one
with $M=0$ in \eqref{PhiAlpha}, and a linear one in $\mu$ with $M=1$. Both
functionals have the same form of \eqref{Phi} with $N_1=3$ and $N_2=8$. The
trained functionals are represented in Fig.~\ref{FigTwo} in terms of the
density profiles minimising their respective cDFTs given in \eqref{OM}. The
cDFT minimisation is done for a variety of pores and chemical potentials, all
of which are chosen outside of the training dataset. The top and bottom plots
in (a)--(c) correspond to the $\mu$-independent model and the linear model,
respectively. Dotted curves show the MAP estimators. The uncertainty of the
inferred $F[\rho]$ is illustrated by plotting 400 samples from the posterior
(grey). The exact $\rho_{X\mid\mu}(x)$ is superimposed in red and
demonstrates a good agreement of the trained functionals with the ground
truth.

As we saw earlier in Fig.~\ref{FigOne}, the trained $F[\rho]$ generalises
well with $L$. Once again, this shows that the inference is consistent with
\eqref{OM}. At large $L$ the bulk fluid densities are again properly
captured, as shown in Fig.~\ref{FigTwo}(c). Our results show that both
trained functionals generalise well to out-of-sample $\mu$, but reveal
interesting and subtle differences. The linear $\mu$-model shows
significantly more confidence in its predictions than the $\mu$-independent
model. This is revealed by the fact that predictive posterior samples of
$\rho(x)$ in the top panel form a much narrower band around their respective
MAP estimators than the bottom panels. A more subtle difference concerns the
two possibilities for the test $\mu$-points: either the chosen $\mu$
extrapolates from the training set [Figs.~\ref{FigTwo} (a) and (b)] or
interpolates it [Fig.~\ref{FigTwo}(c)]. Evidently, during extrapolation
$\mu$-independent model gives slightly better MAP estimators than the linear
model, in spite of the fact that the former has higher uncertainty. Moreover,
when $\mu$ interpolates the training set, the difference between the MAP
estimators nearly vanishes. We can attribute the higher certainty of the
linear model to its higher flexibility in fitting the dependencies. At the
same time, the slightly worse accuracy of the MAP estimator from the more
complex linear model, observed during extrapolation, suggests over-fitting.
The actual $\Phi(n_0, \eta)$ is independent of $\mu$, so artificially
relaxing the $\mu$-dependence may fit the training set with more certainty,
but sacrifices generalisation.

\section{Inferring the canonical functional}
So far we have been considering the grand-canonical ensemble of HR. In other
words, our system was open, with the number of particles fluctuating around a
mean $\langle N_\mu\rangle$, determined by the chemical potential $\mu$. The
exact cDFT of a HR fluid is known only for that case. However, there are many
important problems, where one needs the free-energy functional $F_N[\rho]$ of
a system with a fixed number of particles $N$, i.e. a canonical ensemble
cDFT. For example, in non-equilibrium statistical mechanics $F_N[\rho]$
enters the Fokker-Planck equation for the time-dependent probability density
\cite{LutskoAdv.Chem.Phys.2010, tevrugtAiP2020}. When $N$ is large, the grand-canonical and
canonical ensembles are indistinguishable. But in small systems the
fluctuations of $N$ may be significant, leading to large differences between
the ensembles \cite{LebowitzPercusEtAlPhysRev1967}. Analytically derived
approximate $F_N[\rho]$ requires to solve systems of coupled integral
equations but it is not practical due to its complexity
\cite{WhiteGonzalezEtAlPhysRevLett2000}. Here we statistically infer a simple
and robust approximate $F_N[\rho]$ from particle data.

We simulate $N$ HR by removing the particle insertion-deletion steps from the
grand-canonical simulation, as described in Sec.~Methods. In the canonical
ensemble, the direct problem of statistical mechanics from Sec.~\ref{SecDP}
formally looks the same. We still need to minimise $\Omega[\rho]$ in
\eqref{OM}, with $F[\rho]\equiv F_N[\rho]$. There is, however, one important
distinction. In the canonical ensemble, knowing the system's partition
function is equivalent to knowing the minimal $F_N[\rho]$ and not
$\Omega[\rho]$. Consequently, $F_N[\rho]$ is a function of $N$, and
$\Omega[\rho]$ is simply a Lagrangian of the constrained minimisation
problem. Additionally, $\mu$ is no longer a thermodynamic field, but is
simply a Lagrange multiplier. We can now safely proceed with the inference in
Eqs. (\ref{EL}) and (\ref{Phi}), where we set $\langle N_\mu\rangle\equiv N$.
The training set $\mathcal{D}_N$ contains positions of $N$ particles in $K$
different pores:
\begin{equation}
\label{CanonicalData}
\mathcal{D}_N=\{L_n, \{\left(y_1,\dots y_N\right)_i\}_{i=1}^{M_n}\}_{n=1}^{K}.
\end{equation}

\begin{figure}
\centering
\includegraphics[scale=1]{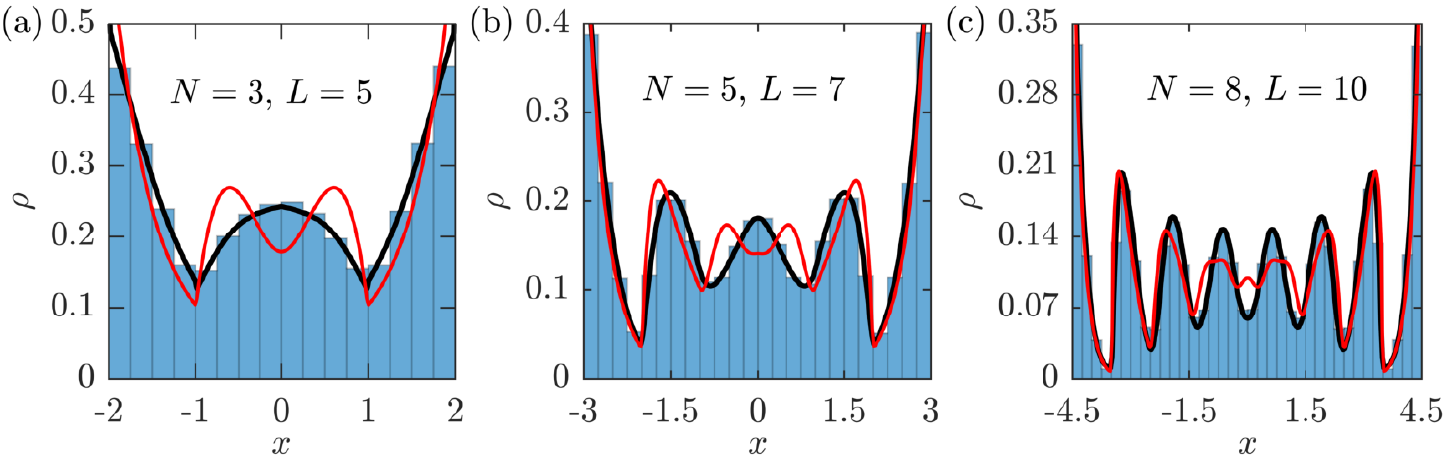}
\caption{Out-of-sample performance of three different inferred canonical cDFT functionals $F_N[\rho]$, for $N$ specified in (a)-(c).
In each case the same model for $F_N[\rho]$ is used, given by \eqref{Phi} with $N_1=N_2=4$. Each $F_N[\rho]$ was trained on a dataset in
\eqref{CanonicalData} with $K=6$, $M_n =10^4$, and $L_n$ equispaced in $[L-2,L+2]$ with step 0.8, where $L$ is specified in the figures.
Black curves show $\rho(x)$ minimising the MAP estimators for the inferred $F_N[\rho]$. The ground truth is represented by the histograms of
simulated particle coordinates at the same $L$ and $N$. Also showing the exact grand-canonical $\rho_{X\mid\mu}(x)$, computed at $\langle N_\mu \rangle=N$ (red).
Observe the excellent agreement of the inferred canonical $F_N[\rho]$ with the ground truth and the break-down of the grand-canonical description of the same system.}
\label{FigThree}
\end{figure}

In Fig.~\ref{FigThree} we plot the inference results for three different
systems with small $N$. For each $L$ in (a)--(c) we trained $F_N[\rho]$ on 6
equispaced $L_n$, chosen between $L-2$ and $L+2$ with a step of 0.8, so that
the test $L$ shown in the figures are not in the training sets. In each case
we trained on $M_n = 10^4$ particle coordinates. Black curves show the
densities minimising the MAP estimators for $F_N[\rho]$, with the ground
truth expressed by the histograms of the particle coordinates. Observe the
remarkable agreement of the inferred DFT with the histograms. To highlight
the difference between the ensembles in each case, we superimpose the exact
grand-canonical $\rho_{X\mid\mu}(x)$, computed at $\mu$, such that $\langle
N_\mu \rangle=N$. We notice that the ensembles differ the most in the pore
centres, where $\rho_{X\mid\mu}(x)$ predicts local extrema. With further
increase of the system size, the difference between the ensembles vanishes as
expected.

\section{Data efficiency}
To assess the data efficiency and accuracy of our method we compare it to a
baseline black-box distributional model. Here we ignore the fact that our
method yields a full cDFT functional of the underlying system and simply
infer the distribution of HR. Consider the following mixture model of
Gaussian radial distribution functions (RBF):
\begin{equation}
\label{StatModel}
\rho(x\mid\mu)=\sum_{i=1}^{N_f}\alpha_i(\mu) \exp{\left(-(x-p_i)^2/w_i^2(\mu)\right)},
\end{equation}
where $\alpha_i(\mu) \geq 0$ for all $i$ and $\sum^{N_f}_j \alpha_j(\mu) =
1$. The Gaussian means $p_i$ are fixed to be equispaced inside the
computational domain to speed-up the training, but we assume $\mu$-dependence
of the remaining free parameters in \eqref{StatModel}. As before, we consider
fixed-$\mu$ and variable-$\mu$ settings, with the respective likelihoods
given in Eqs.~\ref{likelihoodMu} and \ref{likelihood}. For the fixed-$\mu$
model, we place a Gaussian prior on $w_i$ and a rectified prior distribution
with mean 1 and variance 0.1 on $\alpha_i$, forcing  $\alpha_i$ to be
non-negative. For the variable-$\mu$ model, we treat $w_i$ as quadratic
polynomials in $\mu$, and $\alpha_i$ -- as an exponentiated quadratic
polynomial in $\mu$. We then place Gaussian priors with zero means and
variances 0.1 on the polynomial coefficients.

\begin{figure}
\centering
\includegraphics[scale=1]{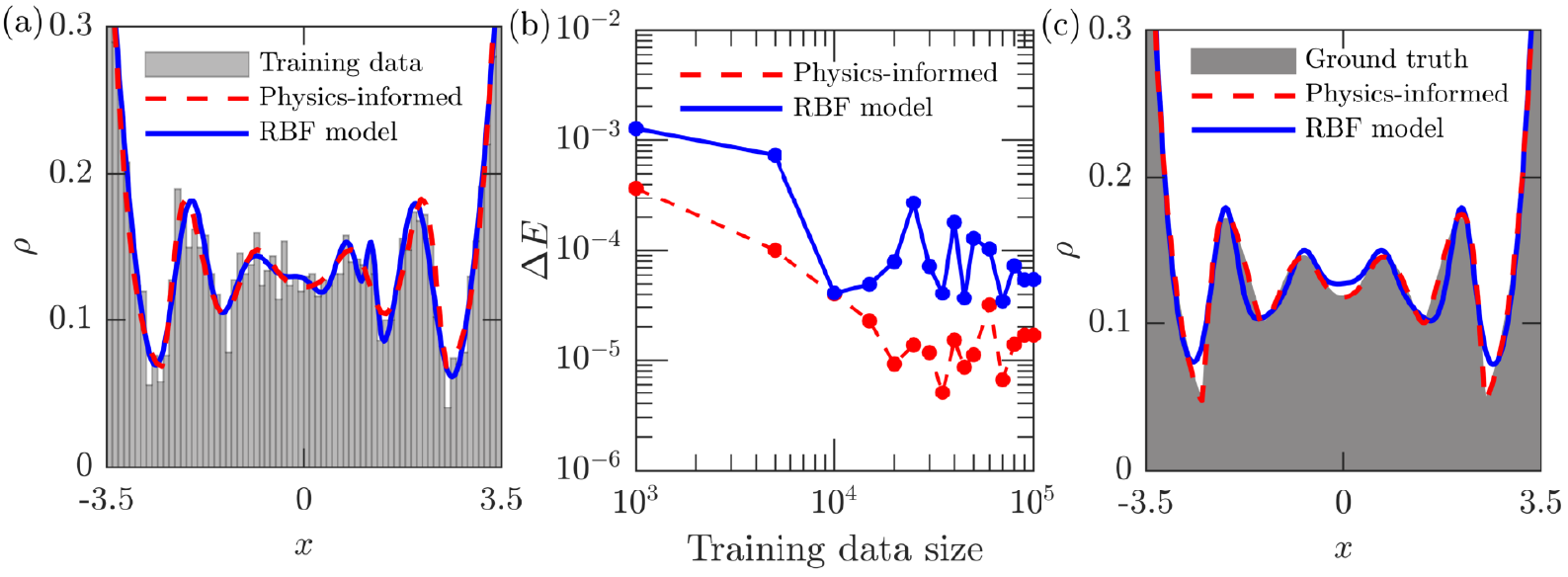}
\caption{Comparison between the physics-informed and black-box approaches to inference. Panels (a) and (b) show fixed-$\mu$ inference, and (c)
shows variable-$\mu$ inference. (a) DFT minimiser of the physics-informed model from \eqref{Phi} with $N_1=N_2=6$ (dashed red), and the black-box model from \eqref{StatModel}
with $N_f=21$ (blue). Both models are trained at $\mu=3$ and $L=8$, on the dataset in \eqref{DataMu} of size $M=5\times10^3$, represented by the histogram.
(b) Energy distance \cite{szekelyJoSPaI2013} to ground-truth as a function of the training data size $M$. (c) Physics-informed model with $N_1=N_2=6$ (red dashed)
and RBF model with $N_f = 10$ (blue). Both are trained at $L=8$ on the dataset in \eqref{Data} with $K=6$, $\mu=2.7, 2.8, 2.9, 3.1, 3.2$ and $3.3$, and $M_i=3\times10^4$;
the ground truth is shown in grey.}
\label{FigFour}
\end{figure}

Figure~\ref{FigFour} represents a comparison between the black-box and
physics-informed approaches. In Fig.~\ref{FigFour}(a) we superimpose two MAP
estimators in the fixed-$\mu$ setting: the RBF distribution (blue) and the
DFT minimiser of the MAP functional (red dashed). Both models were trained on
the same small dataset, represented by the histogram. As we saw earlier with
similar examples, the physics-informed model performs very well in low-data
regimes. In fact, the physics-informed $\rho(x)$ visually coincides with the
ground truth $\rho_{X\mid\mu}(x)$ everywhere, and to keep the figure simple,
we omitted the plot of $\rho_{X\mid\mu}(x)$. The quality of the RBF model is
comparatively worse. There is simply not enough training data to produce an
equally good black-box representation of $\rho_{X\mid\mu}(x)$. This is
revealed by the lack of symmetry. Increasing the data size will improve the
quality of the black-box model. We quantify this in Fig.~\ref{FigFour}(b) by
computing the \emph{energy distance} \cite{szekelyJoSPaI2013}, $\Delta E$,
between $\rho_{X\mid\mu}(x)$ and the MAP-estimators of the inference models
as a function of the training data size. The physics-informed model remains
at least an order of magnitude closer to the ground truth than the black-box
model.

In Fig.~\ref{FigFour}(c) we compare the two approaches in the variable-$\mu$
setting. This time we superimpose the MAP estimators obtained in the regime
of large training data. The data is represented by the histogram, which here
visually coincides with $\rho_{X\mid\mu}(x)$. Again, the physics-informed
model performs remarkably and is visually indistinguishable from the
histogram. The black-box model seems to capture all the essential features of
$\rho_{X\mid\mu}(x)$, but is still inferior to the physics-informed model in
accuracy. A larger RBF basis may improve the representation quality in this
case. The physics-informed approach yields a much more stable model and
requires far fewer training $\mu$-points to achieve good representation. When
extrapolating from the training set over $\mu$, both approaches may struggle
for $\mu$-points far from the training set. Even when the inference becomes
inaccurate, the physics-informed model would still yield symmetric
distributions which satisfy statistical-mechanical sum-rules
\cite{LutskoAdv.Chem.Phys.2010}. Lastly, by construction the physics-informed
model generalises with $L$. We obviously cannot expect this from the
black-box model. The dependence on $L$ must be explicitly built into
\eqref{StatModel}, and then even more data, spanning different $L$, will be
needed for training. In the end, the cost of training a black-box model may
be several orders of magnitude higher than training a physics-informed model.

\section{Conclusion}
In the traditional sense, physical modeling is often associated with analytic
derivations, followed by computation and validation against experimental
data. On the other hand, modern statistical inference offers means to
accomplish similar goals numerically, whilst staying in touch with the data
at all stages of the modelling. Here we focused on the synthesis of both
paradigms. We developed a powerful data-driven, physics-constrained approach
for obtaining humanly interpretable free-energy functionals from small
amounts of data. Our method is fully Bayesian and is based on uncertainty
propagation through all levels of modelling yielding uncertainty
quantification.

We restricted attention to a system with repulsive interactions. In a broader
context, our approach can be applied to systems with more complex
interactions via coarse-graining. For example, if there are long-range
attractions and the repulsive free energy is obtained via inference, a simple
mean-field term can be added to the functional to account for the
attractions. In principle, coarse graining lets us systematically obtain
different terms of the free-energy functional, corresponding to different
parts of interparticle interactions. The generalisation to higher-dimensional
fluids is conceptually straightforward, and can be implemented by considering
functionals of the same family, $\Phi(\{n_i\})$. However, special care should
be taken to properly train the inference model in regions, where the system
undergoes phase transitions. In such cases, the parametric form of
$\Phi(\{n_i\})$ should allow for singular behaviour.

\begin{acknowledgments}
PY was supported by Wave 1 of The UKRI Strategic Priorities Fund
under the EPSRC Grant EP/T001569/1, particularly the “Digital Twins for
Complex Engineering Systems” theme within that grant, and The Alan Turing
Institute. ABD was supported by the Lloyds Register Foundation Programme on
Data Centric Engineering and by The Alan Turing Institute under the EPSRC
grant [EP/N510129/1]. SK was supported by the Engineering and Physical
Sciences Research Council of the UK via grant No. EP/L020564/1.
\end{acknowledgments}

\section*{Appendix}
\label{Methods}
\subsection*{Simulation algorithm}
Here we provide the algorithms to produce a set of particle coordinates of a HRs of radius $R$ confined inside a pore of width $L$ at chemical potential $\mu$. The steps to simulate  the grand-canonical ensemble are enumerated below. The canonical algorithm for a fixed number of particles inside the same pore is obtained by repeating only steps 1-2 with $N=N_i$.
\begin{enumerate}
\setcounter{enumi}{-1}
\item Starting with a random integer $1\leq N_0\leq L/2R$ (initial number of particles in the pore), do 1--3 in a loop over $i=0, 1, \dots$
\item Randomly draw $N_i+1$ non-negative real numbers with the sum equal to $(L-2RN_i)$. These give lengths of $N_i-1$ particle-particle gaps and 2 particle-wall gaps;
\item Compute coordinates $\mathcal{Y}_i=(y_1,\dots y_{N_i})$ of $N_i$ particles from the gaps of step 1;
\item Obtain the new number of particles $N_{i+1}\in\{N_i-1, N_{i}, N_i+1\}$ by attempting particle insertion/deletion with probabilities $P_{\text{del}}/P_{\text{ins}}$;
\begin{align}
& P_{\text{del}}=\{N_i\exp{(-\mu)}/(L/2R) \text{ if } N_i>1 \text{ and } 0 \text{ otherwise,}\}\notag\\
& P_{\text{ins}}=\{(L/2R)\exp{(\mu)}/(N_i+1) \text{ if } N_i+1<L/2R \text{ and } 0 \text{ otherwise}\}
\end{align}
\end{enumerate}

To build a set of $M$ particle coordinates in the grand-canonical ensemble, we run the steps 1--3 for approximately $ML/R$ iterations, to obtain the cumulative flattened data set $\left(\mathcal{Y}_1, \mathcal{Y}_2, \dots\right)$. Then we uniformly thin it to reduce correlations between $y_i$, keeping $M$ particle coordinates $\{y_i\}_{i=1}^{M}$, and compute the expected number of particles in the pore:
\begin{equation}
\label{Nmean}
\langle N_\mu\rangle=\frac{1}{M_\mu}\sum_{i=1}^{M_\mu} N_i.
\end{equation}

\subsection*{Log-posterior and its gradient}
At a fixed $\mu$, the HMC sampler for posterior $P(Q)$ is implemented with the following expressions for log-posterior and its gradient:
\begin{align}
&\log{P}(Q)=\log{\mathcal{N}(Q\mid \bar{Q},\Sigma_Q)}+\sum\limits_{i=1}^{M_\mu}\log{\frac{\rho(y_i\mid Q)}{\langle N_\mu\rangle}},\label{logPmu}\\
&\nabla_Q\log{P(Q)} = -\Sigma_Q^{-1}\left(Q-\bar{Q}\right)+\sum\limits_{i=1}^{M_\mu} \nabla_Q\rho(x) \biggr\rvert_{x = y_i}\frac{1}{\rho(y_i\mid Q)}.\label{gradLogPmu}
\end{align}

\bibliography{references5}

\end{document}